\documentclass[11pt]{article} 
\usepackage{epsfig} 
\usepackage{cite}
\def\e{\epsilon}

\def\ln{\hbox{ln}}

\newcommand{\be}{\begin{equation}}
\newcommand{\eps}{\epsilon}
\newcommand{\dps}{\displaystyle}
\newcommand{\ee}{\end{equation}}
\newcommand{\bea}{\begin{eqnarray}}
\newcommand{\eea}{\end{eqnarray}}

\newcommand{\loopint}[1]{\int \!\!\! \frac{d^D #1}{\left(2\pi\right)^D}\!}
\newcommand{\ESGamma}{S_{\Gamma}}

\newcommand{\lk}{\left(}
\newcommand{\rk}{\right)}
\newcommand{\lek}{\left[}
\newcommand{\rek}{\right]}
\newcommand{\lp}{\left.}
\newcommand{\rp}{\right.}
\newcommand{\nnb}{\nonumber}

\newcommand{\MeijerG}[7]{G^{#1}_{#2} \bigg( #3 \left|
\begin{array}{c}
\left\{ #4 \right\} \, , \, \left\{ #5 \right\} \\
\left\{ #6 \right\} \, , \, \left\{ #7 \right\}
\end{array}
\rk}
\newcommand{\MB}[2]{\hs{-12} \int\limits_{\hs{15}_{ #1 -i \,
\infty}}^{\hs{15}^{ #1 +i\, \infty}} \hs{-15} \frac{d #2}{2\pi i}}
\newcommand{\pFq}[5]{\, \! _{#1} F_{#2}( #3 \, ; \, #4 \, ; \, #5 )}

\newcommand{\hs}[1]{\hspace*{#1 pt}}
\newcommand{\vs}[1]{\vspace*{#1 pt}}

\begin{document} 
\begin{titlepage} 
\vspace*{-1cm} 
\begin{flushright} 
ZU-TH 29/07\\
PITHA-07/17\\
Edinburgh 2007/42\\
SLAC-PUB-13015\\
November 2007
\end{flushright} 
\vskip 3.5cm 

\begin{center} 
{\Large\bf Master Integrals for Fermionic Contributions \\ to Massless Three-Loop Form Factors}
\vskip 1.cm 
{\large G.~Heinrich$^{a}$}, {\large T.~Huber$^{b,c}$} and {\large D.~Ma{\^{\i}}tre$^{d}$}
\vskip .7cm 
{\it $^a$School of Physics, The University of Edinburgh, Edinburgh EH9 3JZ, Scotland, UK}
\vskip .7cm
{\it $^b$Institut f\"ur Theoretische Physik, Universit\"at Z\"urich,
Winterthurerstrasse 190,\\ CH-8057 Z\"urich, Switzerland}
\vskip .7cm
{\it $^c$Institut f\"ur Theoretische Physik E, RWTH Aachen\\D-52056 Aachen, Germany}
\vskip .7cm
{\it $^d$Stanford Linear Accelerator Center, Stanford University, Stanford, CA 94309, USA} 
\end{center} 
\vskip 1cm 

\begin{abstract}

In this letter we continue the calculation of master integrals for massless three-loop form factors by giving analytical results for
those integrals which are relevant for the fermionic contributions proportional to $N_F^2$, $N_F \cdot N$, and $N_F/N$. 
Working in dimensional regularisation, we express one of the integrals 
in a closed form which is exact to all orders in
$\e$, containing $\Gamma$-functions and hypergeometric functions of unit argument. 
In all other cases we derive multiple  Mellin-Barnes representations from 
which the coefficients of the Laurent expansion in $\e$ are extracted in
an analytical form. To obtain the finite part of the three-loop quark 
and gluon form factors, all coefficients through transcendentality 
six in the Riemann $\zeta$-function have to be
included.
\end{abstract}
\vfill 
\end{titlepage} 
\newpage 

\section{Introduction}
The quark form factor $\gamma^\ast \to q \bar q$ and gluon form factor  
$H \to gg$ (effective coupling) are the simplest processes containing 
infrared divergences at  
higher orders in massless quantum field theory, and therefore appear in a 
large variety of physical applications. They can for instance be used to 
predict the infrared pole structure of multi-leg amplitudes at a given 
order~\cite{Magnea:1990zb,catani,sterman} and to extract resummation 
coefficients~\cite{Magnea:2000ss,moch1}, 
and they are needed for the purely virtual corrections to a number of collider 
reactions (Drell-Yan process, Higgs production and decay, DIS).

At the  two-loop level, corrections to the massless quark~\cite{vanneerven} and 
gluon~\cite{harlander,ravindran} form
factors were computed in dimensional 
regularisation with $D=4-2\e$ to order $\e^0$ and subsequently 
extended to all orders in $\e$ in Ref.~\cite{ghm}.
Two-loop corrections to this 
order were also obtained for massive quarks~\cite{breuther}. The
three-loop form factors to order $\e^{-1}$ (and $\e^0$ for 
fermion loop contributions) were computed in~\cite{moch1,moch2}.
One of the main motivations for obtaining analytical results for the form 
factors is the search for a deeper underlying structure of the coefficients, 
as proposed in Ref.~\cite{Bern:2005iz} for planar box amplitudes.

In order to calculate  the quark and gluon form factors 
at higher orders in perturbation theory, the amplitudes 
are reduced, by means of algebraic reduction procedures~\cite{chet,laporta,air,gr}, to a 
small set of master integrals. At the three-loop level, the master integrals for massless
form factors were identified in Ref.~\cite{cedricpaper} and results for certain 
subsets are available in the
literature~\cite{chet,bekavac,mincer,cedricpaper}. 
Among the three-loop master integrals, the genuine three-loop vertex functions 
 are the most challenging ones from a computational point of view. 
They correspond to two-particle cuts of the 
master integrals for massless four-loop off-shell propagator integrals~\cite{baikov},
which have been used in the calculation of the scalar $R$-ratio~\cite{bck}. 
The derivation of the three-loop vertex 
integrals is of comparable complexity to the four-loop propagator 
integrals.

Working in dimensional regularisation and expanding the master integrals in a 
Laurent series in $\e$,  
the finite part of the three-loop form factors requires the extraction 
of all coefficients through transcendentality six, i.e.\ 
coefficients containing terms up to $\pi^6$ or $\zeta_3^2$. 
Note that the power of $\e$ coming with coefficients of transcendentality six 
in the Laurent expansion is not always the same in the different master integrals: 
Transcendentality six can appear in the coefficients of the $\e^0$-, $\e$- or $\e^2$-terms 
in the  Laurent series. If it appears in the $\e^k$-term, this indicates that
a prefactor $\sim 1/\e^k$ will come from the reduction to master integrals, such that 
an expansion up to  transcendentality six of the master integrals will 
always be required.

Those genuine three-loop vertex functions which contain one-loop or two-loop 
propagator insertions were already given in Ref.~\cite{cedricpaper}. 
The purpose of 
the present letter is to extend this calculation to all three-loop master 
integrals which have less than nine propagators. 
Each topology contains only one master 
integral, which is chosen to be the scalar integral, with 
no loop momenta in the numerator and  with all propagators 
raised to unit power. It turns out that this subset of three-loop master integrals is sufficient in order to obtain the aforementioned
fermion loop contributions within a Feynman diagrammatic approach~\cite{gehrmannprivate}. 
At this point we would like to point out an 
error in Ref.~\cite{cedricpaper}, namely the basis of three-loop master integrals given there
is too large, since certain two-particle cuts of four-loop 
propagator topologies \cite{baikov} yield topologically identical 
three-loop vertex topologies. Consequently
$A_{8,1} = A_{8,2} \equiv A_8$ and $A_{9,2} = A_{9,3}$.
The corrected set of three-loop vertex integrals is
given in Fig.~\ref{fig:3loopdiagrams}.

This letter is organised as follows. Computational methods to obtain analytical and numerical results of the 
three-loop vertex integrals with up to eight
propagators are described in Section~\ref{sec:mi}, 
and the analytical results for them are listed in 
Section~\ref{sec:results}. Section~\ref{sec:conc} contains our conclusions and an outlook.

\section{Master integrals: Classification and computational methods}
\label{sec:mi}
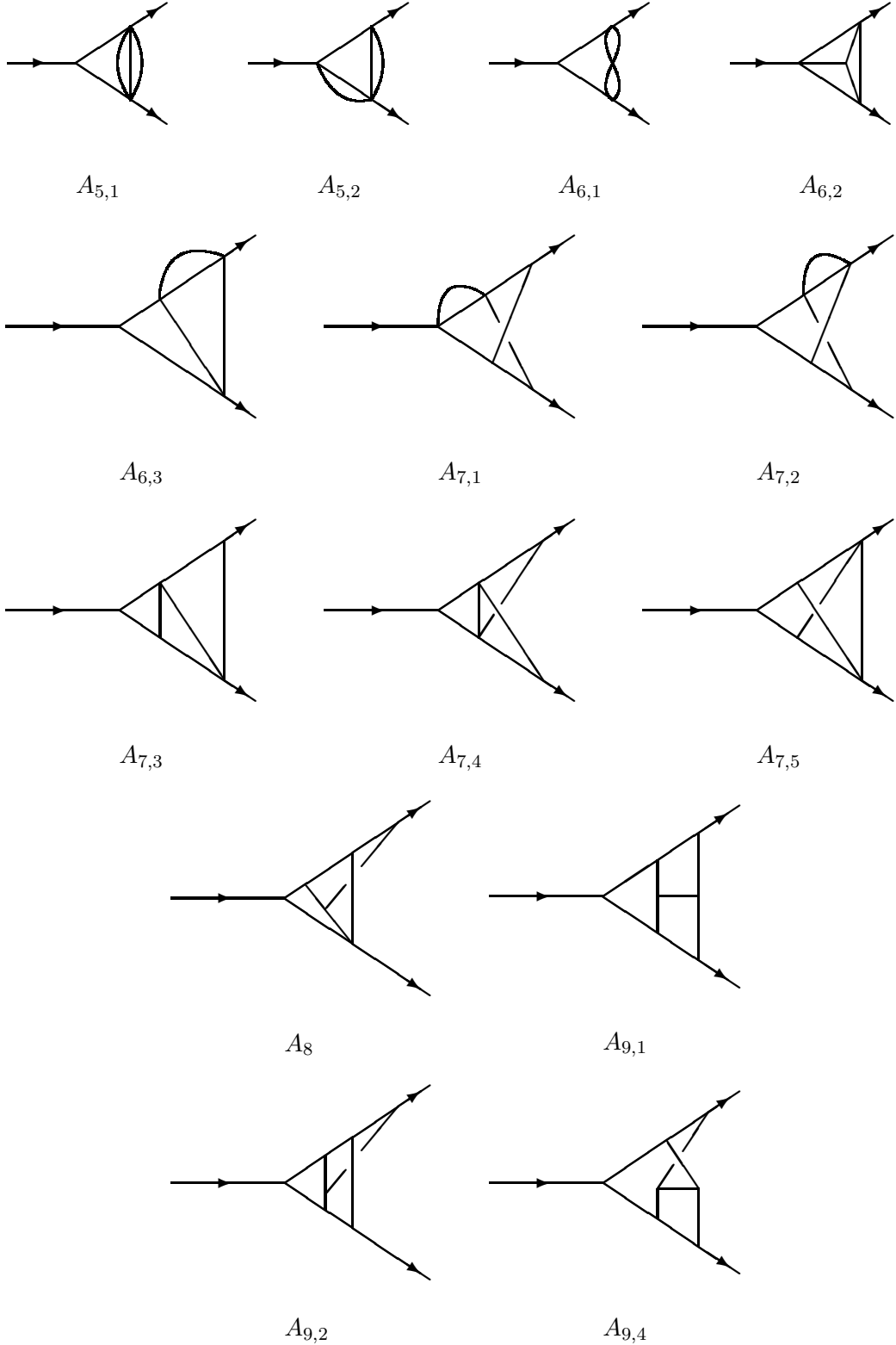
\begin{figure}[p]
    \vs{5}
    \hs{31}
  \begin{tabular}{cccc}
  \parbox{3.3cm}{
\begin{picture}(0,0)
\thicklines
\put(-30,0){\vector(1,0){17}}
\put(-13,0){\line(1,0){13}}
\put(0,0){\line(3,2){40}}
\put(0,0){\line(3,-2){40}}
\put(0,0){\vector(3,2){37}}
\put(0,0){\vector(3,-2){37}}
\put(24,-16){\line(0,1){32}}
\qbezier[80](24,-15.9)(34,0)(24,16)
\qbezier[80](24,-16.1)(12.5,0)(24,16)
\end{picture}\\ \vs{30} \\ $A_{5,1}$}
  & 
\parbox{3.3cm}{
\begin{picture}(0,0)
\thicklines
\put(-30,0){\vector(1,0){17}}
\put(-13,0){\line(1,0){13}}
\put(0,0){\line(3,2){40}}
\put(0,0){\line(3,-2){40}}
\put(0,0){\vector(3,2){37}}
\put(0,0){\vector(3,-2){37}}
\put(24,-16){\line(0,1){32}}
\qbezier[80](24,-16)(34,0)(24,16)
\qbezier[80](0,0)(8,-20)(24,-16)
\end{picture}\\ \vs{30} \\ $A_{5,2}$}
  & 
\parbox{3.3cm}{
\begin{picture}(0,0)
\thicklines
\put(-30,0){\vector(1,0){17}}
\put(-13,0){\line(1,0){13}}
\put(0,0){\line(3,2){40}}
\put(0,0){\line(3,-2){40}}
\put(0,0){\vector(3,2){37}}
\put(0,0){\vector(3,-2){37}}
\qbezier[60](24,16.3)(18,12)(24,0)
\qbezier[60](24,-16.3)(18,-12)(24,0)
\qbezier[60](24,16.3)(30,12)(24,0)
\qbezier[60](24,-16.3)(30,-12)(24,0)
\end{picture}\\ \vs{30} \\ $A_{6,1}$}
  &
\parbox{3.3cm}{
\begin{picture}(0,0)
\thicklines
\put(-30,0){\vector(1,0){17}}
\put(-13,0){\line(1,0){13}}
\put(0,0){\line(3,2){40}}
\put(0,0){\line(3,-2){40}}
\put(0,0){\vector(3,2){37}}
\put(0,0){\vector(3,-2){37}}
\put(27,-18){\line(0,1){36}}
\put(26.8,-17.8){\line(-1,3){5.9}}
\put(26.8,17.8){\line(-1,-3){5.9}}
\put(0,0){\line(1,0){20.8}}
\end{picture}\\ \vs{30} \\ $A_{6,2}$} \\ \\ \\ \\ \\
  \end{tabular}
  \hs{50}
  \begin{tabular}{ccc}
  \parbox{4.5cm}{
\begin{picture}(0,0)
\thicklines
\put(-50,0){\vector(1,0){27}}
\put(-23,0){\line(1,0){23}}
\put(0,0){\line(3,2){60}}
\put(0,0){\line(3,-2){60}}
\put(0,0){\vector(3,2){57}}
\put(0,0){\vector(3,-2){57}}
\put(46.2,-30.8){\line(0,1){61.6}}
\put(18,12){\line(2,-3){28.7}}
\qbezier[80](17.6,12.9)(20,40)(46,31)
\end{picture}\\ \vs{40} \\ $A_{6,3}$}
 &
\parbox{4.5cm}{
\begin{picture}(0,0)
\thicklines
\put(-50,0){\vector(1,0){27}}
\put(-23,0){\line(1,0){23}}
\put(0,0){\line(3,2){60}}
\put(0,0){\line(3,-2){60}}
\put(0,0){\vector(3,2){55}}
\put(0,0){\vector(3,-2){55}}
\put(24,-16){\line(2,5){17.4}}
\put(42,-28){\line(-1,2){10.5}}
\put(27,2){\line(-1,2){6}}
\qbezier[60](0,0)(0,26)(20.7,14)
\end{picture}\\ \vs{40} \\ $A_{7,1}$}
  &   
  \parbox{4.5cm}{
\begin{picture}(0,0)
\thicklines
\put(-50,0){\vector(1,0){27}}
\put(-23,0){\line(1,0){23}}
\put(0,0){\line(3,2){60}}
\put(0,0){\line(3,-2){60}}
\put(0,0){\vector(3,2){55}}
\put(0,0){\vector(3,-2){55}}
\put(24,-16){\line(2,5){17.4}}
\put(42,-28){\line(-1,2){10.5}}
\put(27,2){\line(-1,2){6}}
\qbezier[60](20.7,14)(20,40)(41,27.7)
\end{picture}\\ \vs{40} \\ $A_{7,2}$}    \\ \\ \\ \\ \\
\parbox{4.5cm}{
\begin{picture}(0,0)
\thicklines
\put(-50,0){\vector(1,0){27}}
\put(-23,0){\line(1,0){23}}
\put(0,0){\line(3,2){60}}
\put(0,0){\line(3,-2){60}}
\put(0,0){\vector(3,2){57}}
\put(0,0){\vector(3,-2){57}}
\put(46.2,-30.8){\line(0,1){61.6}}
\put(18,12){\line(2,-3){28.7}}
\put(18,-12){\line(0,1){24}}
\end{picture}\\ \vs{40} \\ $A_{7,3}$}
 &
 \parbox{4.5cm}{
\begin{picture}(0,0)
\thicklines
\put(-50,0){\vector(1,0){27}}
\put(-23,0){\line(1,0){23}}
\put(0,0){\line(3,2){60}}
\put(0,0){\line(3,-2){60}}
\put(0,0){\vector(3,2){57}}
\put(0,0){\vector(3,-2){57}}
\put(18,-12){\line(0,1){24}}
\put(18,-12){\line(2,3){6.7}}
\put(28,3){\line(2,3){18.5}}
\put(18,12){\line(2,-3){28.7}}
\end{picture}\\ \vs{40} \\ $A_{7,4}$}
&
\parbox{4.5cm}{
\begin{picture}(0,0)
\thicklines
\put(-50,0){\vector(1,0){27}}
\put(-23,0){\line(1,0){23}}
\put(0,0){\line(3,2){60}}
\put(0,0){\line(3,-2){60}}
\put(0,0){\vector(3,2){57}}
\put(0,0){\vector(3,-2){57}}
\put(46.2,-30.8){\line(0,1){61.6}}
\put(18,-12){\line(2,3){6.7}}
\put(28,3){\line(2,3){18.5}}
\put(18,12){\line(2,-3){28.7}}
\end{picture}\\ \vs{40} \\ $A_{7,5}$} \\ \\ \\ \\ \\
\end{tabular}
\begin{tabular}{cc}
\hs{125}\parbox{4.5cm}{
\begin{picture}(0,0)
\thicklines
\put(-50,0){\vector(1,0){27}}
\put(-23,0){\line(1,0){23}}
\put(0,0){\line(3,2){64}}
\put(0,0){\line(3,-2){64}}
\put(0,0){\vector(3,2){60}}
\put(0,0){\vector(3,-2){60}}
\put(30,-20){\line(0,1){40}}
\put(30,-20){\line(-4,5){20.7}}
\put(18,-5){\line(5,6){9}}
\put(34,14.2){\line(5,6){15.5}}
\end{picture}\\ \vs{40} \\ $A_{8}$}
  &   
\parbox{4.5cm}{
\begin{picture}(0,0)
\thicklines
\put(-50,0){\vector(1,0){27}}
\put(-23,0){\line(1,0){23}}
\put(0,0){\line(3,2){60}}
\put(0,0){\line(3,-2){60}}
\put(0,0){\vector(3,2){55}}
\put(0,0){\vector(3,-2){55}}
\put(24,-16){\line(0,1){32}}
\put(42,-28){\line(0,1){56}}
\put(24,0){\line(1,0){18}}
\end{picture}\\ \vs{40} \\ $A_{9,1}$} \\ \\ \\ \\ \\
\end{tabular}\\
\begin{tabular}{cc}
\hs{125}\parbox{4.5cm}{
\begin{picture}(0,0)
\thicklines
\put(-50,0){\vector(1,0){27}}
\put(-23,0){\line(1,0){23}}
\put(0,0){\line(3,2){64}}
\put(0,0){\line(3,-2){64}}
\put(0,0){\vector(3,2){60}}
\put(0,0){\vector(3,-2){60}}
\put(18,-12){\line(0,1){24}}
\put(30,-20){\line(0,1){40}}
\put(18,-5){\line(5,6){9}}
\put(34,14.2){\line(5,6){15.5}}
\end{picture}\\ \vs{40} \\ $A_{9,2}$}
&
 \parbox{4.5cm}{
\begin{picture}(0,0)
\thicklines
\put(-50,0){\vector(1,0){27}}
\put(-23,0){\line(1,0){23}}
\put(0,0){\line(3,2){60}}
\put(0,0){\line(3,-2){60}}
\put(0,0){\vector(3,2){56}}
\put(0,0){\vector(3,-2){56}}
\put(24,-16){\line(0,1){13.4}}
\put(42,-28){\line(0,1){25.6}}
\put(23.8,-2.6){\line(1,0){18.4}}
\put(42,-2.5){\line(-2,3){14}}
\put(24.1,-2.6){\line(2,3){7}}
\put(35.1,13.9){\line(2,3){11.3}}
\end{picture}\\ \vs{40} \\ $A_{9,4}$} \\ \\ \\
  \end{tabular}
    \vs{-20} 
  \caption{Three-loop master integrals with massless propagators. The incoming momentum is $q=p_1+p_2$. Outgoing lines are considered
  on-shell and massless, i.e. $p_1^2=p_2^2=0$.  \label{fig:3loopdiagrams}}
  \end{figure}

Vertex integrals with one off-shell and two on-shell legs
and massless propagators depend only on one kinematic scale: the mass $q^2$
of the off-shell leg. The dependence on this scale is given by the 
mass dimension of the integral, such that the coefficients of the  
Laurent expansion in the dimensional regularisation parameter $\e$ are real constants 
 which are in general of 
increasing transcendentality in the Riemann $\zeta$-function, where the degree of 
transcendentality ($DT$) is defined by
\bea
DT(r)&=&0\quad\mbox{ for rational } r\nonumber\\
DT(\pi^k)&=&DT(\zeta(k))=k\nonumber\\
DT(x\cdot y)&=&DT(x)+DT(y)\;.
\eea
At the three-loop level the quark form factor depends -- like the process 
$e^+ e^- \to 3$\,jets at NNLO -- on the following seven colour
structures~\cite{threejetnnlo,gehrmannprivate}
\be
N^2\, , \qquad N^0 \, , \qquad 1/N^2\, , \qquad N_F \cdot N\, , \qquad N_F/N\, , \qquad N_F^2\, ,\qquad  N_{F,\gamma}  \, ,
\ee
where the last colour factor stems from topologies in which the external gauge boson couples to a closed fermion loop.
The three
terms containing $N_F$ are referred to as {\textit{fermionic corrections}}. They have been derived in 
Refs.~\cite{moch1,moch2} from the 
behaviour of the three-loop deep inelastic coefficient 
functions~\cite{moch4}. In the more conventional approach of  
computing  multi-loop Feynman amplitudes the form factors are -- after an algebraic reduction procedure~\cite{chet,laporta,air,gr} --
expressed in terms of a small set of master
integrals. It turns out~\cite{gehrmannprivate} that the master integrals in Fig.~\ref{fig:3loopdiagrams} 
with at most eight propagators are sufficient in order to obtain
the fermionic corrections to the form factor. The purpose of this
letter is therefore to evaluate these master integrals. Those master integrals in Fig.~\ref{fig:3loopdiagrams} that contain single or
multiple bubble insertions have already been computed in Ref.~\cite{cedricpaper}, the remaining ones with up to eight propagators --
i.e.\ diagrams $A_{6,2}$, $A_{7,3}$, $A_{7,4}$, $A_{7,5}$, and $A_8$ -- 
are subject of the present work. Working in
dimensional regularisation with $D=4-2\e$, we give one of the diagrams ($A_{7,4}$) 
in a closed form which is exact to all orders in
$\e$, containing $\Gamma$-functions and hypergeometric functions of unit argument. In all other cases we derive multiple -- twofold to
fourfold -- Mellin-Barnes representations~\cite{smirnov,Tausk,smirnovbook,Alejandro} from which the coefficients of the Laurent expansion in $\e$ are obtained in
an analytic form. As explained above, all coefficients through transcendentality six in the Riemann $\zeta$-function have to be
included to obtain the finite part of the three-loop form factor.

For many practical applications, and to verify the analytical results, 
it is sufficient to know the numerical 
values of the coefficients in the 
Laurent expansion of the master integrals to some finite order. 
There are several techniques to obtain numerical values for the
coefficients, one of them being the sector decomposition method, which 
is described in detail in Refs.~\cite{gudrun1,gudrun2}.
Using this technique, the Laurent expansions of all 
master integrals relevant to the three-loop form factors can be computed 
to, in principle, any desired order. 
In practice there are of course  limitations, from CPU time for the numerical evaluation 
and from memory for the algebraic part of the sector decomposition 
procedure. 
The eight propagator graph $A_{8}$ is the most complex one
from the sector decomposition point of view,  not only due to the high number of 
propagators, but also because it exhibits spurious linear divergences 
at intermediate stages, which render the subtractions and thus the 
functions to be integrated more complicated.
The computing time for  $A_8$  up to order $\eps$ for a 
numerical precision of 0.1\% is of the order 
of 4 hours on a 3.0 GHz PC. 
For a precision of 1\% the evaluation 
is more than 10 times faster.

Another method of doing numerical cross checks proceeds along the lines of
deriving Mellin-Barnes representations of loop integrals by means of the
package {\tt AMBRE}~\cite{ambre} and subsequently performing the analytical
continuation and numerical evaluation of the obtained expressions with the
package {\tt MB}~\cite{czakonMB}. This procedure allowed us to check
most of the coefficients at the sub-permille level.
\section{Results}
\label{sec:results}
In this section we list the results we obtained for the three-loop master integrals necessary for 
the fermionic corrections to the
three-loop quark form factor, which are  the diagrams $A_{6,2}$, $A_{7,3}$, $A_{7,4}$, $A_{7,5}$ 
and $A_8$ in Figure~\ref{fig:3loopdiagrams}. All other diagrams with
up to eight propagators possess so-called bubble insertions and have already been given in Ref.~\cite{cedricpaper}. 

\vspace*{10pt}

\noindent\textit{Diagram $A_{6,2}$}

\vspace*{10pt}

The first diagram to be considered is $A_{6,2}$.  In Ref.~\cite{Huber:2007dx} 
a representation of this
diagram in terms of a one-dimensional integral over hypergeometric
functions was given. Here we pursue a different strategy and derive a 
twofold Mellin-Barnes representation~\cite{Alejandro,Tausk,smirnovbook} from which the coefficients of the Laurent
series expansion about $\eps=0$ can be computed. We start with
\bea
\dps A_{6,2} &=& \loopint{k}\loopint{l}\loopint{r} \; \; \frac{1}{\lk k+p_1\rk^2 \, \lk k+l-p_2\rk^2 \, \lp l \rp^2 \, \lp r \rp^2 \,
\lk r-k\rk^2 \, \lk r-k-l\rk^2} \; ,
\eea
and assume here and in the following that all propagators contain an infinitesimal $+i\eta$. We then derive the following expression
that contains a triple integral over a Meijer-G function~\cite{thebook,thebook2},
\bea
\dps A_{6,2} &=& -i \, \ESGamma^3 \, \lek -\lp q\rp^2-i \, \eta \rek^{-3\, \eps} \frac{\Gamma^3(1-\eps) \,
\Gamma(3\,\eps)}{\Gamma(1-2\,\eps) \, \Gamma(2-4\,\eps)} \nnb \\
&&\nnb \\
&& \times \, \int\limits_0^1 \! dx \, dy \, dz \; x^{-\eps} \, (1-x)^{-3\,\eps} \, y^{-\eps}\,(1-y)^{-3\,\eps} \,
z^{-2\,\eps}\,(1-z)^{-2\,\eps}\nnb \\
&&\nnb \\
&& \times \,  \MeijerG{32}{33}{x \, z + y \, (1-z)}{-1+4\,\eps,-1+4\,\eps}{3 \,\eps}{-1+3\,\eps,-1+2\,\eps,0}{} \; \; ,
\eea
\begin{equation}
{\rm where } \qquad q^2=(p_1+p_2)^2 \qquad {\rm and } \qquad S_{\Gamma} = \frac{1}{\lk 4\pi\rk^{D/2}\,\Gamma(1-\e)} \,.
\end{equation}
We now make use of the contour integral representation of
the Meijer-G function~\cite{thebook,thebook2}, and subsequently decompose the argument by means of a second Mellin-Barnes
representation. The integrals over $x$, $y$, and $z$ can then be done in terms of $\Gamma$-functions. This leads to the
following twofold Mellin-Barnes representation for $A_{6,2}$,
\bea
\dps A_{6,2} &=& -i \, \ESGamma^3 \, \lek -\lp q\rp^2-i \, \eta \rek^{-3\, \eps} \frac{\Gamma^3(1-\eps) \,
\Gamma(3\,\eps) \, \Gamma^2(1-3\,\eps)}{\Gamma(1-2\,\eps) \, \Gamma(2-4\,\eps)} \MB{c_1}{w_1} \MB{c_2}{w_2} \nnb \\
&&\nnb \\
&& \times \,
\frac{\Gamma(-1+3\,\eps-w_1)\,\Gamma(-1+2\,\eps-w_1)\,\Gamma(2-4\,\eps+w_1)\,\Gamma(-w_2)\,\Gamma(w_2-w_1)}{\Gamma(3\,\eps-w_1) \,
\Gamma(2-4\,\eps+w_2) \, \Gamma(2-4\,\eps+w_1-w_2)}\nnb \\
&&\nnb \\
&& \times \, \Gamma(1-\eps+w_2)\,\Gamma(1-\eps+w_1-w_2)\,\Gamma(1-2\,\eps+w_2)\,\Gamma(1-2\,\eps+w_1-w_2) \;\; . \label{eq:A62}
\eea
In the above equation~(\ref{eq:A62}) the contour integrals in the complex 
plane are along curves which separate left poles of
$\Gamma$-functions from right ones, where ``left poles" are poles stemming from a 
$\Gamma(\ldots +w)$ dependence, while ``right poles" stem from a 
$\Gamma(\ldots -w)$ dependence~\cite{smirnovbook}.
The most convenient choice for these contours are
straight lines parallel to the imaginary axis, 
\textit{i.e.}\ the real parts along the curves are constant.
According to Refs.~\cite{Alejandro,Tausk}, these real parts, together
with the parameter $\eps$, must be chosen in such a way as to have positive arguments in all occurring $\Gamma$-functions in order to
separate left and right poles in the desired way. One verifies easily that
\be
c_1 = - \frac{6}{5} \; , \qquad c_2 = - \frac{1}{2}\; ,\qquad -\frac{1}{15} < \eps < \frac{3}{20}
\ee
is an appropriate choice in Eq.~(\ref{eq:A62}). From the fact that the origin lies within the allowed region for $\eps$, we conclude that
the Mellin-Barnes integration does not produce any poles in $\eps$ in addition to the UV pole that is already present in the prefactor.
Therefore the expansion in $\eps$ commutes with the contour integrations. Proceeding in this way, the Mellin-Barnes integrations can be
done order by order in $\eps$. During this procedure, 
the contours can be closed at infinity to either side of the complex plane, 
and the corresponding residues are then summed with the appropriate global sign. 
Furthermore, the nested sums algorithm~\cite{Vermaseren:1998uu,nestedsums} and the formulas in the
Appendix of Ref.~\cite{smirnovbook} -- Barnes Lemmata and corollaries thereof -- prove extremely useful. The final result for $A_{6,2}$
is
\bea
\dps A_{6,2} &=& i \, \ESGamma^3 \lek -\lp q\rp^2-i \, \eta \rek^{-3\, \eps}\nnb\\
&&\nnb \\
&& \times \, \lek -\frac{2 \, \zeta_3}{\eps} - 18 \, \zeta_3 -
\frac{7\pi^4}{180} + \lk  - 122 \, \zeta_3 - \frac{7\pi^4}{20} + \frac{2\pi^2}{3} \, \zeta_3 - 10 \, \zeta_5\rk \, \eps\rp\nnb \\
&&\nnb \\
&& \hs{18}\lp + \lk -738 \, \zeta_3 - \frac{427\pi^4}{180} + 6 \pi^2 \, \zeta_3 - 90 \, \zeta_5+ \frac{163\pi^6}{7560} + 76 \,
\zeta_3^2\rk \, \eps^2 + {\cal O} (\eps^3)\rek \; \; .
\eea
In Ref.~\cite{Huber:2007dx}, two more orders of the $\e$-expansion can be found.

\vspace*{10pt}

\noindent\textit{Diagram $A_{7,3}$}

\vspace*{10pt}

We now turn our attention to the integral
$A_{7,3}$. This integral will be represented, similarly to the integral $A_{6,2}$, 
as a multiple Mellin-Barnes integral:
\bea
\dps A_{7,3} &=& \loopint{k}\loopint{l}\loopint{r} \; \; \frac{1}{\lp k\rp^2 \, \lk k+q\rk^2  \,
\lk l-k-p_2\rk^2 \,\lk l-p_2\rk^2 \, \lk r+l\rk^2\, \lp r \rp^2 \, \lk r-p_1\rk^2} \nnb \\
&&\nnb\\
&=& i \, \ESGamma^3 \, \lek -\lp q\rp^2-i \, \eta \rek^{-1-3\, \eps} \frac{\Gamma^4(1-\eps) \,
\Gamma(-\eps)}{\Gamma(1-2\,\eps) \, \Gamma(1-3\,\eps)} \MB{c_1}{w_1} \MB{c_2}{w_2} \MB{c_3}{w_3} \;
\frac{\Gamma(-w_1)}{\Gamma(1-w_1)}\nnb\\
&&\nnb \\
&& \times \,
\frac{\Gamma(-3\,\eps-w_3) \,\Gamma(1+2\,\eps+w_1+w_2)\,\Gamma(1+w_1+w_2)\,\Gamma(-2\,\eps-w_2)
\,\Gamma(-\eps-w_1)}{\Gamma(1-3\,\eps-w_3) \, \Gamma(2-2\,\eps+w_1+w_2)}\nnb \\
&&\nnb \\
&& \times \, \Gamma(-w_3)\,\Gamma(\eps-w_1-w_2+w_3)\,\Gamma(1-\eps+w_2)\,\Gamma(1+w_3) \, \Gamma(-\eps+w_1-w_3) \;\; . \label{eq:A73}
\eea
The contour integrals are again along straight lines in the complex plane parallel to the imaginary axis, and as before we must 
choose the real parts of the integration variables such as to have positive arguments in all occurring $\Gamma$-functions. This is
achieved by choosing
\be\label{eq:ciforA73}
c_1 = - \frac{3}{20} \; , \qquad c_2 = - \frac{3}{5}\; ,\qquad c_3 = - \frac{1}{2}\; ,\qquad -\frac{1}{8} < \eps < \frac{3}{20}.
\ee
As it was the case for $A_{6,2}$, we have the origin within the allowed region for $\eps$ and therefore the Mellin-Barnes integration does
not give rise to any additional poles in $\eps$, the only pole 
of the integral being the infrared pole that is already present in the prefactor in
Eq.~(\ref{eq:A73}). We can thus again perform the contour integrations order by order in $\eps$. Since the leading coefficient turns out
to have already transcendentality five, we only need to compute the first two terms in the expansion. They are given by
\be
\dps  A_{7,3} = i \, \ESGamma^3 \lek -\lp q\rp^2-i \, \eta \rek^{-1-3\, \eps} \,\lek \lk -\frac{\pi^2\,\zeta_3}{6}-10
\,\zeta_5\rk\frac{1}{\eps} - \frac{119 \,\pi^6}{2160} - \frac{31}{2} \, \zeta_3^2 + {\cal O} (\eps)\rek  . 
\ee
During the evaluation of this integral we could not proceed until the end by merely applying Barnes Lemmata and corollaries thereof, but
 had to apply auxiliary integral representations of hypergeometric functions at intermediate steps. The benefit of this procedure is
that it enables us to perform all Mellin-Barnes integrations, at the cost of introducing additional parameters over which we
subsequently have to integrate. However, the integrations over these auxiliary parameters 
can be done in terms of logarithms and
(harmonic) polylogarithms. Furthermore, we made extensive use of the package {\tt HPL}~\cite{HPL,HPL2} and of an algorithm based on the
nested sums approach~\cite{Vermaseren:1998uu,nestedsums}. See Appendix~\ref{app:detailsMB} for more details on this point.

\vspace*{10pt}

\noindent\textit{Diagram $A_{7,4}$}

\vspace*{10pt}

The next diagram we consider is $A_{7,4}$. At first glance it looks quite difficult since it lacks both a bubble insertion and a
planar topology. However, it turns out to be simpler than the planar diagram $A_{7,3}$, and it can even be displayed in a closed form.
The main reason for this is the fact that at the outer vertices of both outgoing lines only three lines meet, and hence the introduction
of Feynman parameters allows for the combination of propagators that differ only by a 
light-like momentum. This property is
absent in both $A_{6,2}$ and $A_{7,3}$, both of which did not reveal a closed form but only a multiple Mellin-Barnes representation. For
$A_{7,4}$, we find
\bea
\dps A_{7,4} &=& \loopint{k}\loopint{l}\loopint{r} \; \; \frac{1}{\lp k\rp^2 \, \lk k-q\rk^2  \,
\lk r+l-k\rk^2 \,\lp l\rp^2 \, \lk l-p_1\rk^2\, \lp r \rp^2 \, \lk r-p_2\rk^2} \nnb \\
&&\nnb\\
&=& i \, \ESGamma^3 \, \lek -\lp q\rp^2-i \, \eta \rek^{-1-3\, \eps} \cdot 2 \cdot \Gamma^4(1-\eps) \, \Gamma^2(-\eps) \nnb \\
&&\nnb\\
&& \times \bigg[ \frac{\Gamma(1-\eps)\,\Gamma(3\,\eps)}{\lk1-3\,\eps\rk^2\,\Gamma(2-4\,\eps)}
\; \pFq{4}{3}{1,1-\eps,1-3\,\eps,2-6\,\eps}{2-3\,\eps,2-3\,\eps,2-4\,\eps}{1}\nnb \\
&&\nnb\\
&&\hs{20} -\frac{\Gamma(1-3\,\eps)\,\Gamma(2-3\,\eps)\, \Gamma(3\,\eps)\,\Gamma(1+2\,\eps)}{\Gamma(2-\eps)\,\Gamma(2-6\,\eps)} \nnb \\
&&\nnb\\
&&\hs{35} \times \, \pFq{4}{3}{1,1,1+2\,\eps,2-3\,\eps}{2,2,2-\eps}{1}\nnb \\
&&\nnb\\
&&\hs{20} +\frac{\Gamma^2(1-3\,\eps)\,\Gamma(1+2\,\eps)\,\Gamma(1+3\,\eps)}{\Gamma(2-\eps)\,\Gamma(2-6\,\eps)} \nnb \\
&&\nnb\\
&&\hs{35} \times \, \pFq{4}{3}{1,1,1+2\,\eps,1+3\,\eps}{2,2,2-\eps}{1} \bigg]\nnb \\
&&\nnb \\
&=& i \, \ESGamma^3 \, \lek -\lp q\rp^2-i \, \eta \rek^{-1-3\, \eps}\nnb \\
&&\nnb\\
&& \times \bigg[ \frac{6\,\zeta_3}{\eps ^2} + \lk \frac{11\,\pi^4}{90} + 36\,\zeta_3 \rk\frac{1}{\eps} + 
 \lk\frac{11\,{\pi }^4}{15} + 216\,\zeta_3 - 2\,{\pi }^2\,\zeta_3 + 46\,\zeta_5\rk \nnb \\
&&\nnb \\
&&\hs{15}  + \lk \frac{22\,{\pi }^4}{5} - \frac{19\,{\pi }^6}{270} +  1296\,\zeta_3 - 12\,{\pi }^2\,\zeta_3 - 
    282\,\zeta_3^2 + 276\,\zeta_5 \rk \, \eps  + {\cal O}(\eps^2) \bigg] ,  \label{eq:A74}
\eea
where the expansion in $\e$ was done by means of the {\tt Mathematica}~\cite{Mathematica} package {\tt
HypExp}~\cite{hypexp,Huber:2007dx}.

\vspace*{10pt}

\noindent\textit{Diagram $A_{7,5}$}

\vspace*{10pt}

We now consider diagram $A_{7,5}$ for which we derive the following fourfold Mellin-Barnes representation.
\bea
\dps A_{7,5} &=& \loopint{k}\loopint{l}\loopint{r} \; \; \frac{1}{\lp k\rp^2 \, \lk k+q\rk^2  \,
\lk k+r\rk^2 \,\lk l-p_2\rk^2 \, \lk r-l\rk^2\, \lp r \rp^2 \, \lk k+l+p_1\rk^2} \nnb \\
&&\nnb\\
&=& i \, \ESGamma^3 \, \lek -\lp q\rp^2-i \, \eta \rek^{-1-3\, \eps} \frac{\Gamma^3(1-\eps) \,}
{\Gamma(1-2\,\eps) \, \Gamma(1-4\,\eps)} \MB{c_1}{w_1} \MB{c_2}{w_2} \MB{c_3}{w_3} \MB{c_4}{w_4}\nnb\\
&&\nnb \\
&& \,
\frac{\Gamma(w_4-w_1) \,\Gamma(1+w_3) \,\Gamma(-3\eps-w_3) \,\Gamma(1-2\eps+w_1+w_2-w_4) \,\Gamma(-w_3)\,\Gamma(-w_4)}
{\Gamma(1-w_1+w_3+w_4)}\nnb \\
&&\nnb \\
&& \times \, \frac{\Gamma(1+3\eps+w_3+w_4)\,\Gamma(1+\eps+w_1+w_2)\,\Gamma(1+w_1+w_2)\,\Gamma(-\eps-w_1) \, \Gamma(-\eps-w_2)}
{\Gamma(2+\eps+w_1+w_2) \, \Gamma(2-2\eps+w_1+w_2)}\nnb \\
&&\nnb \\
&& \times \, \frac{\Gamma(1-\eps+w_2) \,\Gamma(1+w_3) \,\Gamma(1-\eps+w_1) \,\Gamma(\eps-w_1-w_2+w_3+w_4)
\,\Gamma(w_4-w_2)}{\Gamma(1-w_2+w_3+w_4)} \;\; .\nnb\\ \label{eq:A75}
\eea
Like in the previous cases, the Mellin-Barnes integral does not generate poles in $\eps$, 
so we can therefore interchange the expansion
in $\eps$ with the contour integrations. We choose
\be
c_1 = - \frac{1}{5} \; , \qquad c_2 = - \frac{1}{4}\; ,\qquad c_3 = - \frac{1}{7}\; ,\qquad c_4 = - \frac{1}{11}.
\ee
As before in the case of $A_{7,3}$ we can not proceed until the end by merely applying Barnes Lemmata and corollaries thereof, but
again have to apply auxiliary integral and series representations at intermediate steps, this time even for a larger class of functions
than before. Besides hypergeometric functions, these are mainly logarithms and (harmonic) polylogarithms as well as $\psi$-functions
with
\bea
\dps \psi^{(0)}(z) &=& \frac{d}{dz} \ln\lek\Gamma(z)\rek \; , \nnb\\
&&\nnb\\
\dps \psi^{(k)}(z) &=& \frac{d}{dz} \psi^{(k-1)}(z) \quad \mbox{for} \quad k = 1,2,\ldots \qquad .
\eea
The sums and integrals over the auxiliary parameters are then performed by means of the same techniques as before. Some details can again be found in Appendix A. The final result for
$A_{7,5}$ reads
\be
\dps  A_{7,5} = i \, \ESGamma^3 \lek -\lp q\rp^2-i \, \eta \rek^{-1-3\, \eps} \,\lek  2\pi^2\,\zeta_3+10
\,\zeta_5 + \lk 12\pi^2\,\zeta_3+60
\,\zeta_5+\frac{11 \pi^6}{162} + 18 \, \zeta_3^2\rk \eps + {\cal O} (\eps^2)\rek  . 
\ee

\vspace*{10pt}

\noindent\textit{Diagram $A_8$}

\vspace*{10pt}

The last diagram we consider is $A_8$ which can also be displayed as a fourfold Mellin-Barnes integral.
\bea
\dps A_8 &=& \loopint{k}\loopint{l}\loopint{r} \; \; \frac{1}{\lk k+p_1\rk^2 \, \lk k+r\rk^2  \,
\lk k+r+q\rk^2 \,\lk l-k\rk^2 \, \lk l+r\rk^2\, \lp l \rp^2 \, \lp r \rp^2 \, \lk l+p_1\rk^2} \nnb \\
&&\nnb\\
&=& -i \, \ESGamma^3 \, \lek -\lp q\rp^2-i \, \eta \rek^{-2-3\, \eps} \frac{\Gamma^3(1-\eps) \,\Gamma(-1-3\eps)}
{\Gamma(-2\,\eps) \, \Gamma(-4\,\eps)} \MB{c_1}{w_1} \MB{c_2}{w_2} \MB{c_3}{w_3} \MB{c_4}{w_4}\nnb\\
&&\nnb \\
&& \,
\frac{\Gamma(1+w_3) \,\Gamma(1+w_4) \,\Gamma(w_4-w_2) \,\Gamma(w_3-w_1)\, \Gamma(-w_4) \,\Gamma(-w_3)\,\Gamma(2+w_1+w_2)}
{\Gamma(2+w_3+w_4) \,\Gamma(1+w_4-w_2) \,\Gamma(1+w_3-w_1)}\nnb \\
&&\nnb \\
&& \times \, \frac{\Gamma(2+\eps+w_1+w_2)\,\Gamma(1-\eps+w_1)\,\Gamma(1-\eps+w_2)\,\Gamma(-1-\eps-w_1) \, \Gamma(-1-\eps-w_2)}
{\Gamma(2-2\eps+w_1+w_2) \, \Gamma(3+\eps+w_1+w_2)}\nnb \\
&&\nnb \\
&& \times  \,\Gamma(2+3\eps+w_3+w_4) \,\Gamma(1-2\eps+w_1+w_2-w_3-w_4)\,\Gamma(\eps-w_1-w_2+w_3+w_4)
\;\; .\nnb\\ \label{eq:A8}
\eea
This time the Mellin-Barnes integral does indeed generate poles in $\eps$. We choose~\cite{czakonMB}
\be
c_1 = - \frac{7}{8} \; , \qquad c_2 = - \frac{19}{24}\; ,\qquad c_3 = - \frac{13}{24}\; ,\qquad c_4 = - \frac{25}{48} \; ,\qquad
-\frac{5}{16} < \eps < - \frac{5}{24}
\ee
in order to separate left poles of $\Gamma$-functions from right ones, and subsequently perform the analytic continuation to
$\eps=0$~\cite{czakonMB}.
This generates four kernels, one four-dimensional one, two three-dimensional ones, and one two-dimensional one. We arrive at the final result
\bea
\dps  A_8 &=& i \, \ESGamma^3 \lek -\lp q\rp^2-i \, \eta \rek^{-2-3\, \eps} \, \bigg[ \frac{8\zeta_3}{3\eps^2} + \lk\frac{5\pi^4}{27}-8
\zeta_3\rk\frac{1}{\eps} + 24\zeta_3-\frac{5\pi^4}{9}-\frac{52}{9} \, \pi^2\zeta_3 +\frac{352}{3} \,\zeta_5 \nnb \\
&&\nnb \\
&&\hs{15}  + \lk - 72\zeta_3+\frac{5\pi^4}{3}+\frac{52}{3} \, \pi^2\zeta_3 -352 \,\zeta_5
+\frac{1709\pi^6}{8505}-\frac{332}{3}\zeta_3^2\rk \, \eps  + {\cal O}(\eps^2) \bigg] .  \label{eq:A8erg}
\eea
Despite the fact that this integral has one more propagator compared to $A_{7,5}$ it was much simpler to evaluate than the former one, and we did not have to
introduce any auxiliary parameters in integral or series representations, but could proceed until the end by the same techniques as described in $A_{6,2}$.
The reason for the simplicity of $A_8$ is again that it possesses -- contrary to $A_{7,5}$ -- an outgoing line with an outer vertex where only three lines meet. 

\section{Conclusions and Outlook}\label{sec:conc}
In this letter we have evaluated those master integrals for massless three-loop form factors 
which are necessary for the
calculation of the fermionic corrections to the quark form factor. We obtained analytical results for all coefficients through
transcendentality six in the Riemann $\zeta$-function, as required 
to obtain the finite part of the form factor at the three-loop level.
For the integral $A_{7,4}$  we could obtain a 
representation which is valid to all orders in $\e$, 
in terms of hypergeometric functions of unit argument.
For the other integrals, we derived multiple Mellin-Barnes representations 
from which we extracted all necessary coefficients  order by order in $\e$.

The only missing pieces to complete the set of master integrals for massless three-loop form factors are therefore the three
diagrams in Figure~\ref{fig:3loopdiagrams} which have nine propagators. 
It turns out that each of them can be expressed in terms of a
sixfold Mellin-Barnes representation~\cite{nineprop} which gives rise to ${\cal O}(100)$ single terms upon performing the
analytical continuation to $\e=0$ by means of the package {\tt MB}~\cite{czakonMB}. Although the number of single integrals is quite
large due to the extraction of high poles in $\e$, and the evaluation is not completely automated at certain
stages, the analytical results for these integrals are within reach~\cite{nineprop}.

\section*{Acknowledgements}
We would like to thank Thomas Gehrmann and Beat T{\"o}dtli for communicating to us part of the structure of the three-loop form factor
prior to publication. We thank Alejandro Daleo and David Kosower for useful discussions and for performing some cross checks. We
would also like to thank Martin Beneke and Pietro Falgari for useful discussions. T.H. was supported
by  the Swiss National Science Foundation (SNF) and by Deutsche Forschungsgemeinschaft,
SFB/TR 9 ``Computergest\"{u}tzte Theoretische Teilchenphysik''. 
The work of D.M. was supported by the Swiss National Science Foundation (SNF) 
under contract  PBZH2-117028 and by the US Department of Energy under contract DE-AC02-76SF00515.
The work of G.H. was supported by the UK Science and Technology Facilities Council. 
The authors would like to thank the Institute for Theoretical Physics at the 
University of Z\"urich, where most of this work was done.

\appendix
\section{Some technical details on performing MB integrals} \label{app:detailsMB}

This appendix is devoted to some details about those terms in diagrams $A_{7,3}$ and $A_{7,5}$ which require the introduction of
auxiliary parameters in addition to the MB variables. Consider the term
\be\label{eq:MBsampleint}
\MB{c_2}{w_2} \MB{c_3}{w_3} \; \left(-\frac{3\pi^2 \csc^2(\pi w_2) \Gamma(w_2+1) \Gamma(1-w_3) \Gamma(-w_3) \Gamma^3(w_3)}{(w_2+1) \Gamma(w_2+w_3+2)}\right)
\ee
which appears in the computation of $A_{7,3}$ at the stage where two MB integrations are left (the first MB integration of this integral
can be done by Barnes Lemma and corollaries thereof, see~\cite{smirnovbook}). The $c_i$ are as in Eq.~(\ref{eq:ciforA73}), and we have converted certain combinations of $\Gamma$-functions
to {\tt Csc}.

We perform the integration over $w_3$ by closing the contour to the right. After summing all residues, one obtains
\bea\label{eq:MBsampleintnoch1}
\MB{c_2}{w_2} &&\Big[\frac{\pi^2 \csc^2(\pi w_2) \big[\psi^{(0)}(w_2+2)\big]^3}{2\, (w_2+1)^2}+
\frac{5 \pi^4 \csc^2(\pi w_2) \psi^{(0)}(w_2+2)}{4\, (w_2+1)^2}\nnb \\
&&+\frac{3 \pi^2 \csc^2(\pi w_2) \big[\psi^{(0)}(w_2+2)+\gamma_E\big]\big[\gamma_E \psi^{(0)}(w_2+2)+\psi^{(1)}(w_2+2)\big]}{2 \, (w_2+1)^2} \nnb \\
&&-\frac{\pi^2 \csc^2(\pi w_2) \psi^{(2)}(w_2+2)}{(w_2+1)^2}+\frac{\pi^2 \csc^2(\pi w_2) \big[2\gamma_E^3+5\gamma_E
\pi^2+4\zeta_3\big]}{4\, (w_2+1)^2}\nnb \\
&& +\frac{3\pi^2 \csc^2(\pi w_2) \pFq{4}{3}{1,1,1,1}{2,2,w_2+3}{1}}{(w_2+1)^2 (w_2+2)}\Big] \; ,
\eea
where $\gamma_E$ is Euler's constant. The integration of all but the last term in the above equation can be carried out with the standard technique of the nested sums algorithm~\cite{Vermaseren:1998uu,nestedsums}.
In the last term we write the hypergeometric function as
\be
\int\limits_0^1 \! dt \, \frac{(w_2+2) (1-t)^{w_2+1} Li_2(t)}{t} \; .
\ee
The integration over $w_2$ can now be done in this term as well. In the end, the integration over the auxiliary parameter $t$ can be carried out
by means of the package {\tt HPL}~\cite{HPL,HPL2}. This yields $9\zeta_3^2+349\pi^6/15120$ for the expression in Eq.~(\ref{eq:MBsampleint}). An alternative approach would be to
replace in Eq.~(\ref{eq:MBsampleint})
\be
\frac{\Gamma(w_2+1)}{\Gamma(w_2+w_3+2)} = \frac{1}{\Gamma(w_3+1)} \, \int\limits_0^1 \! dt \; t^{w_2} \, (1-t)^{w_3}
\ee
with the benefit of having factorized the integrand in $w_2$ and $w_3$. The integrations over $w_2$ and $w_3$ can now be carried out by summation of appropriate residues, followed by integration over $t$.

As far as $A_{7,5}$ is concerned, we consider the finite piece of Eq.~(\ref{eq:A75}) where $\eps$ is set to zero. We perform the integration over $w_4$ and close the contour to the right.
The result contains two hypergeometric functions which we decompose as follows
\bea
&&\pFq{4}{3}{-w_1, -w_2, w_3 + 1, -w_1 - w_2 + w_3}{-w_1 - w_2, -w_1 + w_3 + 1, -w_2 + w_3 + 1}{1} = \nnb \\
&&\nnb \\
&&\int_0^1 \!dt_1 \int_0^1 \!dt_2 \int_0^1 \!dt_3 \, 
  \frac{\Gamma(-w_2 + w_3 + 1) \Gamma(-w_1 + w_3 + 1) \Gamma(-w_1 - w_2)}{\Gamma(-w_1)\Gamma(-w_2)\Gamma(-w_1 - w_2 + w_3)\Gamma^2(1 + w_3)\Gamma(-w_3)} \nnb \\
  &&\nnb \\
  && \times \, \frac{t_1^{-w_1 - 1} (1 - t_1)^{w_3} \, t_2^{-w_2 - 1} (1 - t_2)^{w_3} \, t_3^{-w_1 - w_2 + w_3 - 1}(1 - t_3)^{-w_3 - 1}}{(1 - t_1 t_2 t_3)^{w_3 + 1}} \; ,
\eea
\bea
&&\pFq{4}{3}{w_1 + 1, w_2 + 1, w_3 + 1, w_1 + w_2 + w_3 + 2}{w_1 + w_2 + 2, w_1 + w_3 + 2, w_2 + w_3 + 2}{1} = \nnb \\
&&\nnb \\
&&\int_0^1 \!dt_1 \int_0^1 \!dt_2 \int_0^1 \!dt_3 \, 
  \frac{\Gamma(w_1 + w_2 + 2) \Gamma(w_1 + w_3 + 2) \Gamma(w_2 + w_3 + 2)}{\Gamma(w_1+1)\Gamma(w_2+1)\Gamma(w_1 + w_2 + w_3 +2)\Gamma^2(1 + w_3)\Gamma(-w_3)} \nnb \\
  &&\nnb \\
  && \times \, \frac{t_1^{w_1} (1 - t_1)^{w_3} \, t_2^{w_2} (1 - t_2)^{w_3} \, t_3^{w_1 + w_2 + w_3 + 1}(1 - t_3)^{-w_3 - 1}}{(1 - t_1 t_2 t_3)^{w_3 + 1}} \; .
\eea
We then perform the integration over $w_3$, followed by the other two MB integrations. In the end, we carry out the integrations over $t_1$, $t_2$ and $t_3$. We make
simple variable changes in the $t_1$-$t_2$-$t_3$ cube where appropriate.


\begin{thebibliography}{99} 

\bibitem{Magnea:1990zb}
  L.~Magnea and G.~Sterman,
  Phys.\ Rev.\  D {\bf 42} (1990) 4222.
  
\bibitem{catani}
S.\ Catani, Phys.\ Lett.\ B {\bf 427} (1998) 161
[hep-ph/9802439].

\bibitem{sterman}
G.~Sterman and M.E.~Tejeda-Yeomans,
Phys.\ Lett.\ B {\bf 552} (2003) 48
[hep-ph/0210130].
  
\bibitem{Magnea:2000ss}
  L.~Magnea,
  Nucl.\ Phys.\  B {\bf 593} (2001) 269
  [arXiv:hep-ph/0006255].

\bibitem{moch1}
  S.~Moch, J.A.M.~Vermaseren and A.~Vogt,
  JHEP {\bf 0508} (2005) 049
  [hep-ph/0507039].

\bibitem{vanneerven}
G.\ Kramer and B.\ Lampe, Z.\ Phys.\ C {\bf 34} (1987) 497;
 {\bf 42} (1989) 504(E);\\
T.\ Matsuura and W.L.\ van Neerven, Z.\ Phys.\ C {\bf 38} (1988) 623;\\
T.\ Matsuura, S.C.\ van der Maarck and W.L.\ van Neerven,
Nucl.\ Phys.\ B {\bf 319} (1989) 570.

\bibitem{harlander}
R.V.~Harlander,
Phys.\ Lett.\ B {\bf 492} (2000) 74
[hep-ph/0007289].

\bibitem{ravindran}
  V.~Ravindran, J.~Smith and W.L.~van Neerven,
  Nucl.\ Phys.\ B {\bf 704} (2005) 332
  [hep-ph/0408315].

\bibitem{ghm}
T.~Gehrmann, T.~Huber and D.~Ma\^{\i}tre,
  Phys.\ Lett.\ B {\bf 622} (2005) 295
  [hep-ph/0507061].

\bibitem{breuther}
W.~Bernreuther {\it et al.},
  Nucl.\ Phys.\ B {\bf 706} (2005) 245
  [hep-ph/0406046];
  Nucl.\ Phys.\ B {\bf 712} (2005) 229
  [hep-ph/0412259];
  Nucl.\ Phys.\ B {\bf 723} (2005) 91 [hep-ph/0504190].

\bibitem{moch2}
  S.~Moch, J.A.M.~Vermaseren and A.~Vogt,
  Phys.\ Lett.\ B {\bf 625} (2005) 245
  [hep-ph/0508055].

\bibitem{Bern:2005iz}
  Z.~Bern, L.~J.~Dixon and V.~A.~Smirnov,
  Phys.\ Rev.\  D {\bf 72} (2005) 085001
  [arXiv:hep-th/0505205].

\bibitem{chet}
F.V.\ Tkachov, Phys.\ Lett.\ {\bf 100B} (1981) 65;\\
K.G.\ Chetyrkin and F.V.\ Tkachov, Nucl.\ Phys.\ {\bf B192} (1981) 159.

\bibitem{laporta}
S.~Laporta,
Int.\ J.\ Mod.\ Phys.\ A {\bf 15} (2000) 5087
[hep-ph/0102033].

\bibitem{gr}
T.\ Gehrmann and E.\ Remiddi, Nucl.\ Phys.\ B
{\bf 580} (2000) 485 [hep-ph/9912329].

\bibitem{air}
  C.~Anastasiou and A.~Lazopoulos,
  JHEP {\bf 0407} (2004) 046 
  [hep-ph/0404258].

\bibitem{cedricpaper}
  T.~Gehrmann, G.~Heinrich, T.~Huber and C.~Studerus,
  Phys.\ Lett.\  B {\bf 640}, 252 (2006)
  [arXiv:hep-ph/0607185].

\bibitem{bekavac}
  S.~Bekavac,
  Comput.\ Phys.\ Commun.\  {\bf 175} (2006) 180
  [arXiv:hep-ph/0505174].

\bibitem{mincer}
S.G.~Gorishnii, S.A.~Larin, L.R.~Surguladze and F.V.~Tkachov,
Comput.\ Phys.\ Comm.\  {\bf 55} (1989) 381;\\
S.A.~Larin, F.V.~Tkachov and J.A.M.~Vermaseren,
NIKHEF-H-91-18.

\bibitem{baikov}
 P.A.~Baikov,
  Phys.\ Lett.\ B {\bf 634} (2006) 325
  [hep-ph/0507053].
  
\bibitem{bck}
  P.A.~Baikov, K.G.~Chetyrkin and J.H.~K\"uhn,
  Phys.\ Rev.\ Lett.\  {\bf 96} (2006) 012003
  [hep-ph/0511063].

\bibitem{gehrmannprivate}
T.~Gehrmann and B.~T{\"o}dtli, private communication

\bibitem{threejetnnlo}
  A.~Gehrmann-De Ridder, T.~Gehrmann, E.~W.~N.~Glover and G.~Heinrich,
  JHEP {\bf 0711} (2007) 058.
  [arXiv:0710.0346 [hep-ph]].
  
\bibitem{moch4}
J.A.M.~Vermaseren, A.~Vogt and S.~Moch,
  Nucl.\ Phys.\ B {\bf 724} (2005) 3
  [hep-ph/0504242].

\bibitem{smirnov}
  V.A.~Smirnov,
  %
  Phys.\ Lett.\ B {\bf 460} (1999) 397
  [hep-ph/9905323].

\bibitem{Tausk}
 J.B.~Tausk,
  %
  Phys.\ Lett.\ B {\bf 469} (1999) 225
  [hep-ph/9909506].
  
\bibitem{Alejandro}
  C.~Anastasiou and A.~Daleo,
  JHEP {\bf 0610} (2006) 031
  [arXiv:hep-ph/0511176].
  
\bibitem{smirnovbook}
    V.~A.~Smirnov,
   {\it ``Evaluating Feynman integrals"}, 
  Springer Tracts Mod.\ Phys.\  {\bf 211} (2004) 1.
  
\bibitem{gudrun1}
T.~Binoth and G.~Heinrich,
Nucl.\ Phys.\ B {\bf 585} (2000) 741
[hep-ph/0004013].

\bibitem{gudrun2}
T.~Binoth and G.~Heinrich,
Nucl.\ Phys.\ B  {\bf 680} (2004) 375
[hep-ph/0305234].

\bibitem{ambre}
  J.~Gluza, K.~Kajda and T.~Riemann,
  Comput.\ Phys.\ Commun.\  {\bf 177} (2007) 879
  [arXiv:0704.2423 [hep-ph]].

\bibitem{czakonMB}
  M.~Czakon,
  Comput.\ Phys.\ Commun.\  {\bf 175} (2006) 559
  [arXiv:hep-ph/0511200].

\bibitem{Huber:2007dx}
  T.~Huber and D.~Ma\^{\i}tre,
  arXiv:0708.2443 [hep-ph].

\bibitem{thebook}
A. Erd\'elyi~(ed.),
\newblock Higher transcendental functions, Vol. 1, (McGraw-Hill, New York,
  1953) .
  
 \bibitem{thebook2}
A. Erd\'elyi~(ed.),
\newblock Tables of Integral Transforms, Vol. 2, (McGraw-Hill, New York,
  1954) .

\bibitem{Vermaseren:1998uu}
  J.~A.~M.~Vermaseren,
  Int.\ J.\ Mod.\ Phys.\  A {\bf 14} (1999) 2037
  [arXiv:hep-ph/9806280].
  
\bibitem{nestedsums}
  S.~Moch, P.~Uwer and S.~Weinzierl,
  J.\ Math.\ Phys.\  {\bf 43} (2002) 3363
  [arXiv:hep-ph/0110083].

\bibitem{HPL}
  D.~Ma\^{\i}tre,
  Comput.\ Phys.\ Commun.\  {\bf 174} (2006) 222
  [arXiv:hep-ph/0507152].
  
\bibitem{HPL2}
  D.~Ma\^{\i}tre,
  arXiv:hep-ph/0703052.

\bibitem{Mathematica}
{\textit {MATHEMATICA} 5.2}, {C}opyright 2005 by {W}olfram {R}esearch.

\bibitem{hypexp}
  T.~Huber and D.~Ma\^{\i}tre,
  Comput.\ Phys.\ Commun.\  {\bf 175} (2006) 122
  [hep-ph/0507094].
  
\bibitem{nineprop}
 T.~Huber et.~al., work in progress

\end{thebibliography}
\end{document}